\documentclass[aps,pre,reprint,floatfix,unsortedaddress]{revtex4-1}
\pdfoutput=1

\usepackage{amsmath}
\usepackage{amsfonts}
\usepackage{amssymb}
\usepackage[usenames,dvipsnames]{color}
\usepackage[T1]{fontenc}
\usepackage{graphicx}
\usepackage[hidelinks]{hyperref}
\usepackage{ifpdf}
\usepackage{times}
\usepackage[normalem]{ulem}

\newcommand{\thi}{\ensuremath{\theta_i}}
\newcommand{\thr}{\ensuremath{\theta_r}}
\newcommand{\vect}[1]{\mathbf{#1}}
\newcommand{\re}{\operatorname{Re}}
\newcommand{\im}{\operatorname{Im}}

\hypersetup{
  colorlinks   = true, 
  urlcolor     = blue, 
  linkcolor    = blue,
  citecolor   = blue 
}

\begin{document}

\title{Asymptotic dynamics of reflecting spiral waves}

\author{Jacob Langham}
\email[]{J.Langham@warwick.ac.uk}
\affiliation{Mathematics Institute, University of Warwick, Coventry, CV4 7AL,
United Kingdom}
\author{Irina Biktasheva}
\email[]{ivb@liverpool.ac.uk}
\affiliation{Department of Computer Science, University of Liverpool, Liverpool
L69 3BX, United Kingdom}
\author{Dwight Barkley}
\email[]{D.Barkley@warwick.ac.uk}
\affiliation{Mathematics Institute, University of Warwick, Coventry, CV4 7AL,
United Kingdom}

\date{\today}

\begin{abstract} 
Resonantly forced spiral waves in excitable media drift in straight-line paths,
their rotation centers behaving as point-like objects moving along trajectories
with a constant velocity.  Interaction with medium boundaries alters this
velocity and may often result in a reflection of the drift trajectory. Such
reflections have diverse characteristics and are known to be highly nonspecular
in general. In this context we apply the theory of response functions, which via
numerically computable integrals, reduces the reaction-diffusion equations
governing the whole excitable medium to the dynamics of just the rotation center
and rotation phase of a spiral wave.  Spiral reflection trajectories are
computed by this method for both small- and large-core spiral waves in
the Barkley model. Such calculations provide insight into the process of
reflection as well as explanations for differences in trajectories across
parameters, including the effects of incidence angle and forcing amplitude.
Qualitative aspects of these results are preserved far beyond the asymptotic
limit of weak boundary effects and slow resonant drift.
\end{abstract}

\maketitle

\section{Introduction}
In the past decade an intrinsic wave-particle dualism in spiral waves
has been
highlighted~\cite{bikta03,Biktasheva:2006,Biktashev:2010,Biktasheva:2010,Biktashev:2011}.
This invites comparison with a growing number of macroscopic systems in which
waves propagating in a nonlinear medium are associated with some degree of
spatial localization~\cite{Perrard:2014}, including liquid `walker' droplets
bouncing on a vibrated bath~\cite{Couder:2005,Couder:2005_2}, various optical
solitons~\cite{Stegeman:2000,Grelu:2012} and chemical wave
segments~\cite{Sakurai:2002}. Among other common properties, each of these
examples exhibits nonspecular reflections from obstacles or medium
perturbations~\cite{Protiere:2006,Eddi:2009,Shirokoff:2013,Prati:2011,
Steele:2008} and the dynamics involved in the reflection process can be quite
complex~\cite{lang13}.  It is within this context that we have undertaken the
present investigation.


%
Our study focuses on rotating spiral waves in a system with 
excitable dynamics.
First witnessed experimentally in the Belousov-Zhabotinsky
chemical oscillator~\cite{Belousov:1959,Zhabotinsky:1971,Winfree:1972},
they have since been discovered in diverse
biological~\cite{Tomchik:1981,Tyson:1989,Gorelova:1983,Davidenko:1992,Pertsov:1993},
chemical~\cite{Jakubith:1990,Nettesheim:1993,Agladze:2000} and
physical~\cite{Frisch:1994} contexts.
Within two-dimensional homogeneous excitable media, 
spiral waves typically rotate
about an unexcited core of fixed radius and center.  These are so-called
\emph{rigidly rotating} spirals. The rotation frequency is determined solely by
medium properties, while the center of rotation and phase are determined by
initial conditions.
However, applying spatial or temporal perturbations to an otherwise homogeneous
medium can cause the wave pattern to undergo a spatial displacement or
\emph{drift}~\cite{Biktashev:2007,Biktasheva:2010}. By tracking either the local
rotation center, or the closely related wave tip, one may observe interesting
trajectories as drifting spirals move through a medium. 

\begin{figure}[h]
\centering
\includegraphics[width=3.25in]{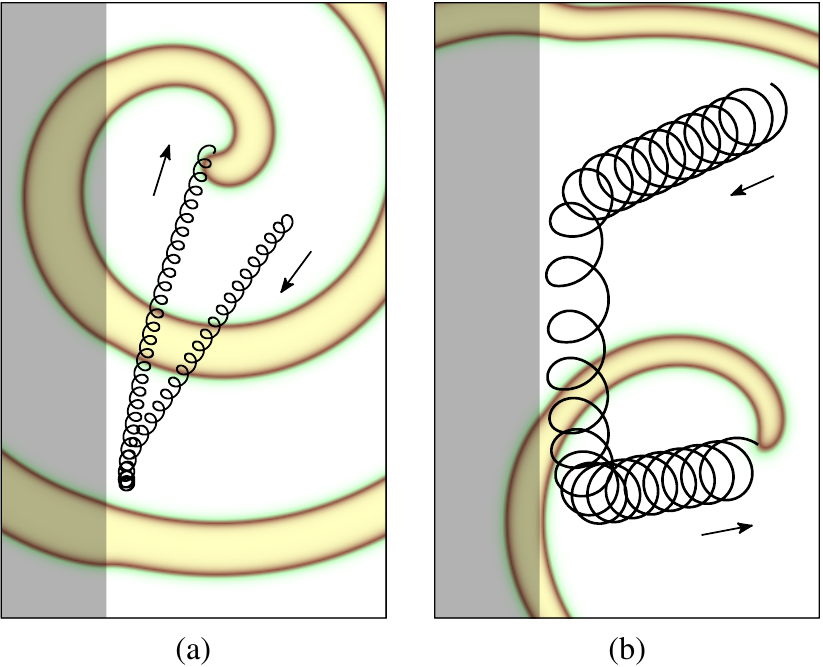}
\caption{
(Color online)
Two examples of resonantly drifting spirals reflecting in the Barkley
model of a generic excitable medium. The trajectories of the wave tips are drawn
in black. Arrows indicate the overall direction of drift. The spiral waves at
the final point in the plotted trajectory are visualized by the $u$-field of the
Barkley model. Both plots use the same length scale.
The boundaries are generated by a step change in medium properties, indicated by
gray shading at the left-hand edges.
(a) A `small-core' spiral wave approaches a boundary and doubles back on itself; its
reflection angle lies on the same side of the boundary normal as its incidence
angle. (b) A `large-core' spiral wave speeds up close to the boundary and travels
alongside it for a short while before reflecting sharply away. 
(The plots were cropped to $25\times 40$ space units from simulations
performed on a $50 \times 50$ square domain, discretized in space with grid
spacing $h = 1/12$ and in time with time step $\triangle t = 2.09 \times
10^{-3}$. The step change was located $12$ space units from the left-hand domain
wall.  Parameters: (a) $a = 0.8, b = 0.05, c = 0.02, \epsilon_s = 0.035,
\epsilon_f = 1.44 \times 10^{-3}, \omega_f = 1.7893$; (b) $a = 0.6, b = 0.07, c
= 0.02, \epsilon_s = 0.035, \epsilon_f = 4.4 \times 10^{-4}, \omega_f = 0.9504$.
Details concerning these parameters and the methods used are given in
Sec.~\ref{sec:methods}.)
}
\label{fig:reflection_examples} 
\end{figure}

A noteworthy case is \emph{resonant}
drift~\cite{Agladze:1987,Davydov:1988,Steinbock:1993,bikt93,Zykov:1994,Mantel:1996,Zhang:2004,Kantrasiri:2005,Ning-Jie:2006,Xu:2012}
in which spatially uniform periodic driving is applied in resonance with the
spiral rotation frequency. In this case the spiral core travels in a straight
line with constant velocity. In a typical experimental domain, such a spiral
will inevitably come close to a boundary, which may lead to a reflection in the
drift trajectory~\cite{bikt93,Olmos:2008,lang13}, as illustrated in
Fig.~\ref{fig:reflection_examples}. 
Reflections are in general nonspecular: the incidence angle rarely equals the
reflection angle.  Furthermore, the character of individual reflection
trajectories depends on the medium in which the wave propagates, the properties
of the boundary and the spiral's resonant drift velocity. 

Numerical simulations of resonantly drifting spiral reflections were undertaken
some time ago by Biktashev and Holden~\cite{bikt93}, who laid the foundations of
the asymptotic approach in a subsequent study~\cite{bikt95}. Their
numerical work has recently been updated with more extensive simulations and the
calculation of a large catalog of reflection trajectories~\cite{lang13}. A key
feature of spiral wave reflections in these two studies is that the angle of
reflection is essentially independent of the angle of incidence for a large
range of incident angles.  Indeed, the reflection angle instead depends more
strongly on the characteristics of the medium than on incident angle. This was
predicted by Biktashev and Holden using an ordinary differential equation (ODE)
model based on the simplifying assumption that the component of the spiral's
drift velocity caused by interaction with the boundary decays exponentially with
distance from the boundary~\cite{bikt93,bikt95}.  However, a more detailed
theoretical treatment is required to fully understand the mechanism behind
spiral reflection. While separate theoretical accounts of both resonant
drift~\cite{bikt93,bikt95,Biktasheva:1999,Biktasheva:2010,Xu:2012} and spatial
medium
inhomogeneities~\cite{ErmakovaPertsov:1986,Aranson-etal:1995,Biktasheva:2000,Xu:2009,Biktasheva:2010}
(which may act as boundaries to drift) already exist, it is the combination and
interaction of these two phenomena which we must consider here.

A good candidate for an updated approach is to use the theory of \emph{response
functions}~\cite{bikt95,Biktasheva:1999,Biktasheva:2000,bikta03,Biktasheva:2006,Biktasheva:2009,Biktasheva:2010}
which has developed and matured in the years since the Biktashev-Holden study.
Response functions are adjoint modes to the neutral symmetry modes of a spiral
which characterize how the position and rotation phase of a spiral
react to asymptotically small perturbations. 
In practical terms, response functions allow us to reduce the partial
differential equations (PDEs) governing the whole medium to the dynamics of just
three real variables---the two spatial coordinates of the wave rotation
center and the rotational phase.

In this paper we bring the reflection of drifting spirals into this asymptotic
framework by considering the superposition of two small perturbations: one
corresponding to resonant forcing generating drift and the other corresponding
to a step change in a medium parameter acting as a boundary to drift.  Previous
studies addressed both effects independently using response
functions~\cite{Biktasheva:1999,Biktasheva:2010}.
While the approach is strictly applicable only in the limit of slow resonant
drift and weak boundary effects, we show that it nevertheless can capture, and
thereby explain, most of the important features of spiral wave reflections
outside of this asymptotic limit.


\section{Theory}\label{sec:theory}
The underlying dynamics of the excitable medium are well described by models in
the class of reaction-diffusion PDEs on the plane:
\begin{equation}
\partial_t \vect{u} = \vect{D}\nabla^2\vect{u} +
\vect{f}(\vect{u},\vect{p})
\label{eq:reactdiff}
\end{equation}
where $\vect{u}(\vect{x},t) \in \mathbb{R}^\ell$ is a vector of $\ell\geq 2$
state variables for the medium, $\vect{f}(\vect{u},\vect{p}) \in
\mathbb{R}^\ell$ describes the excitable dynamics at each point in space
dependent on a vector of $m$ parameters $\vect{p}\in\mathbb{R}^m$ and $\vect{D}
\in \mathbb{R}^{\ell\times\ell}$ is a (symmetric) diffusion matrix.

We are interested in models that admit solutions rotating with angular
frequency $\omega$ about a center point $R = (X,Y)$. That is, rigidly rotating
waves of the form
\begin{equation}
\vect{u} = \vect{U}(\rho,\vartheta + \omega t - \Phi)
\label{eq:rigid}
\end{equation}
where $(\rho,\vartheta)$ are polar coordinates centered at $R$ and $\Phi$ is
the fiducial phase of the spiral at $t=0$.
Note that due to symmetries of the plane, if Eq.~\eqref{eq:reactdiff} admits a
solution of the form in Eq.~\eqref{eq:rigid}, then there are infinitely many
such solutions related by symmetry, and this is captured by the fact that $R$
and $\Phi$ are arbitrary constants.
We refer to $\omega$ as the \emph{natural} frequency since it is an intrinsic
property of the medium, whereas $R$ and $\Phi$ depend on initial data.

Suppose we perturb the medium slightly.  In the limit of weak perturbations,
this induces small shifts in the rotation center $R$ and the phase $\Phi$,
leaving the shape of the spiral otherwise unchanged.  Thus the response of the
spiral to weak perturbations is a trajectory through the space of solutions of
the form Eq.~\eqref{eq:rigid}, where $R$ and $\Phi$ depend on time.

Mathematically, we treat such a perturbation as the addition of a vector
$||\epsilon \vect{h}(\vect{x},t)||\ll 1$ to the right-hand side of
Eq.~\eqref{eq:reactdiff}. It can be shown using perturbation
methods~\cite{bikt95,Biktasheva:1999,Biktasheva:2006} that to first order in
$\epsilon$, the time derivatives of $R(t)$ and $\Phi(t)$ are proportional to
the $L^2$ inner products $\langle\cdot,\cdot\rangle$ of the spiral's response
functions $\vect{W}_{0}$ and $\vect{W}_1$ with the perturbation vector,
averaged over one full rotation period $T=2\pi/\omega$:
\begin{align}
\dot{\Phi}(t) &= \frac{\epsilon}{T} \int_{t-T/2}^{t+T/2}
\left\langle \vect{W}_{0}, \vect{h} \right\rangle
d\tau + O(\epsilon^2)
\label{eq:Phidot_general}\\
\dot{R}(t) &= \frac{\epsilon}{T} \int_{t-T/2}^{t+T/2} e^{i(\Phi-\omega \tau)}
\left\langle \vect{W}_{1}, \vect{h} \right\rangle d\tau +
O(\epsilon^2)
\label{eq:Rdot_general}
\end{align}
where we use the identification $R=(X,Y) \equiv X + iY$.

Technical details can be found in the appendix and
elsewhere~\cite{bikt95,Biktasheva:1999,bikta03,Biktasheva:2006,Biktasheva:2009,Biktasheva:2010},
but the essence of these equations is the following.  The response functions
are adjoint fields corresponding to the symmetries of the reaction-diffusion
system [Eq.~\eqref{eq:reactdiff}]. $\vect{W}_0$ is $\mathbb{R}^\ell$-valued and
corresponds to the presence of rotational symmetry.  One can think of the
perturbation, $\epsilon \vect{h}$, as providing an infinitesimal impulse
$\epsilon\left\langle \vect{W}_{0}, \vect{h} \right\rangle$ along the direction
of the symmetry (phase $\Phi$ in this case), at each time $\tau$.
Equation~\eqref{eq:Phidot_general} captures the effect of all such impulses
over one rotation period to give the rate of change in $\Phi$.

The response function $\vect{W}_1$ is $\mathbb{C}^\ell$-valued and corresponds
to the two translational symmetries.
Here the perturbation at each time $\tau$ provides the spiral with an
infinitesimal impulse in the direction
$\arg{\langle\vect{W}_1,\vect{h}\rangle}$ rotated by $e^{i(\Phi-\omega\tau)}$
due to the underlying natural rotation of the spiral.  These contributions,
averaged over one rotation period, give the drift velocity.
Importantly, a change in $\Phi$ typically implies a change in the direction of
drift.

Response functions have been computed numerically for a variety of spiral
waves in previous studies. For various cases, including that of the spiral waves
we study here, the support of these functions was found to be highly localized
around the spiral rotation
center~\cite{bikta03,Biktasheva:2006,Biktasheva:2010}. Thus, a spiral wave
drifts only in response to perturbations very close to the core. That is, it
behaves as a particle whose position may be identified with the rotation center
$R$.

We are interested in the case where a resonantly forced spiral moves towards,
and reflects from, a boundary in the medium.
This is a combination of two perturbations to the original reaction-diffusion
equations---a homogeneous, time-periodic one that causes resonant drift of the
spiral and a spatial one that imposes a boundary to the drifting spiral.
Let us suppose the resonant forcing can be described by some $\vect{h}_f(t)$.
In practice we will consider the simple case of harmonic forcing of one of the
medium parameters at the natural frequency $\omega$.
Likewise, suppose that the effect of a boundary may be formulated in
$\vect{h}_s(\vect{x})$.  The type of boundary we shall consider is a sharp
interface along the line $x=0$ between two media with different excitability
properties.  Although this is not a physical barrier to wave propagation, a
drifting spiral core may nevertheless reflect from the spatial inhomogeneity;
see Fig.~\ref{fig:reflection_examples} and Ref.~\cite{lang13}. We refer to this
as a \emph{step boundary}. It may be considered as a weak perturbation provided
that the step change in medium parameters is small. In previous studies a
Neumann or `no-flux' boundary was also considered.  While this type of boundary
condition cannot be treated as a weak perturbation, it has previously been
observed that reflections from a step inhomogeneity are qualitatively similar
to the no-flux case~\cite{lang13}.

The total perturbation to the medium can be written as $\vect{h}(\vect{x},t) =
\epsilon_s \vect{h}_s(\vect{x}) + \epsilon_f \vect{h}_f(t)$, where $0 <
\epsilon_s,\epsilon_f \ll 1$ represent the strengths of the respective `step'
and `forcing' perturbations.  One can immediately see from
Eqs.~\eqref{eq:Phidot_general} and~\eqref{eq:Rdot_general} that the effects of
the two perturbations on $\dot{\Phi}$ and $\dot{R}$ are a linear superposition
and may therefore be considered separately. 
It may consequently be shown (see the appendix) that the equations of motion for
the spiral center $R=(X,Y)$ and phase $\Phi$ are of the form
\begin{align}
\dot{X} &= \epsilon_s S_X(X) + \epsilon_f F_X(\Phi) 
\label{eq:Xdot} \\
\dot{Y} &= \epsilon_s S_Y(X) + \epsilon_f F_Y(\Phi)
\label{eq:Ydot} \\
\dot{\Phi} &= \epsilon_s S_\Phi (X) 
\label{eq:Phidot}
\end{align}
where $S_X$, $S_Y$, $S_\Phi$ are contributions due to the step boundary and
$F_X$, $F_Y$ are contributions due to the resonant forcing. These are given by
integrals of the form in Eqs.~\eqref{eq:Phidot_general}
and~\eqref{eq:Rdot_general}. 
While the functions depend in detail on the specific model used and the
particular spiral wave under consideration, their general form, in particular
their respective dependence on $X$ and $\Phi$ as indicated, is independent of
these details.  

Since the step boundary is located along the line $x=0$ in the original PDE,
the dynamics of the spiral depends only on the distance $X$ of the spiral
center from step boundary and does not depend on $Y$. Likewise, since the step
perturbation is time independent, its effect, when averaged over a full spiral
rotation, cannot depend on the spiral's phase $\Phi$.

The form of the functions $F_X$ and $F_Y$ and the role of $\Phi$ are quite
important. In the appendix we show that for sinusoidal resonant forcing
of a medium parameter:
\begin{equation}
F(\Phi) = Ae^{i\Phi}
\label{eq:resforcing}
\end{equation}
where $F \equiv F_X + iF_Y$ and $A$ is a real constant for each model and set
of parameter choices.  
Hence, for a given spiral wave and given forcing amplitude, the drift velocity
due to resonant forcing is, in the asymptotic limit, constant with direction
determined by the phase $\Phi$. This direction of drift can change as a result
of interaction with the boundary, i.e.,\ the function $S_\Phi$, but not due to
periodic forcing alone.

Equations~\eqref{eq:Xdot},~\eqref{eq:Ydot} and~\eqref{eq:Phidot} reduce the
spiral dynamics from a set of nonlinear PDEs to three coupled autonomous
nonlinear ODEs. The functions $S_X$, $S_Y$, $S_\Phi$, $F_X$, and $F_Y$ on the
right-hand sides must in practice be obtained numerically by taking appropriate
inner products with numerically computed response functions.  Nevertheless,
evaluating the right-hand sides and then numerically solving the ODEs can be
done quickly with minimal computational resources.  It is worth noting that the
essential dynamical quantities $X$, $Y$, and $\Phi$ are the same variables that
Biktashev and Holden used in their asymptotic theory of spiral
reflections~\cite{bikt93,bikt95}.
Moreover, we stress that while the variable $\Phi$ was introduced as the phase
of the spiral wave, its role in the reduced system becomes the direction of
drift due to periodic forcing. 


\section{Model and Methods}\label{sec:methods}
The previous discussion of response functions did not depend on any specific
model. Here, we consider spiral wave solutions in the standard Barkley
model~\cite{bark91,bark08}, for which $\ell = 2$:
\begin{align}
& \frac{\partial u}{\partial t} =  
\nabla^2u + \frac{1}{c}u(1-u)\left(u-\frac{v+b}{a}\right), \\
& \frac{\partial v}{\partial t} = u - v.
\end{align}
The two state variables $u(x,y,t)$ and $v(x,y,t)$ capture, respectively, the
excitation and recovery processes of the medium. Parameters $a, b > 0$ control
the threshold for excitation and $0 < c \ll 1$ sets the timescale of the fast
excitation process, relative to recovery. (The parameter $c$ is usually called
$\epsilon$ but we will not use that notation here.)  For fixed parameter $c$
and variable $a,b$, the section of parameter space which admits rigidly
rotating spiral wave solutions is divided roughly into two regimes
distinguished by the size of the rotation core. The reflective properties of
so-called \emph{small-} and \emph{large-core} spirals markedly
differ~\cite{lang13} and we therefore divide our study along these lines.

Throughout our study we have varied the $b$ parameter to create the step
inhomogeneity by considering $b(x) = b_0 + \epsilon_s (H(x) - 1)$, where $H$ is
the Heaviside step function. Resonant forcing has been applied homogeneously by
varying the excitability $c$ as $c(t) = c_0 + \epsilon_f\cos(\omega_f
(t-t_0))$, where $\omega_f$ is the forcing frequency required to obtain
resonant drift and $t_0$ is some initial forcing time (the choice of
which is discussed in the appendix). For our results in
Sec.~\ref{sec:results}, $\omega_f = \omega$. In all numerical simulations, the
values of $\epsilon_s$ and $\epsilon_f$ have been chosen small enough that the
perturbed medium remains in the same parameter regime (of small- or large-core
rigid rotation) as the unperturbed parameters.

The response functions and natural rotation frequencies for various
small- and large-core spirals in the Barkley model were calculated on a polar
grid using the software \textsc{DXSpiral}~\cite{dxspiral}. The numerical methods
are detailed in Ref.~\cite{Biktasheva:2009}. A disk of radius $15$ was used in
the small core with $64$ angular grid points and $1875$ radial grid points. In
the large core the radius size was increased to $20$ and the number of radial
grid points used was $2500$. The resulting response function discretizations
were used to numerically compute the right-hand sides of
Eqs.~\eqref{eq:Xdot},~\eqref{eq:Ydot}, and~\eqref{eq:Phidot} (see the appendix for
the specific integrals), again using \textsc{DXSpiral}. Reflection trajectories
were calculated by timestepping the resulting three dynamical variables from
chosen initial conditions.

Direct numerical simulations of the Barkley model PDEs were also performed for
comparison with the response function predictions. These were computed using the
standard finite-difference techniques described in
Refs.~\cite{bark91,Dowle:1997}. 
These simulations use unusually high precision to ensure that they correctly
capture the spiral rotation frequency~\cite[Sec.\ IV B]{Biktasheva:2010}. (The
simulations involve forcing at the natural frequency, i.e.\ $\omega_f = \omega$,
obtained very accurately from \textsc{DXSpiral}. Small inaccuracies in the
simulations, which would normally be irrelevant, result in artificial frequency
mismatches which then lead to artificially curved trajectories.)  In the small
core (Fig.~\ref{fig:dnsmultiSC}) a $20 \times 20$ square domain was used with
grid spacing $h=0.0125$ and time step $\triangle t = 2.3 \times 10^{-5}$. The
step inhomogeneity was located $5$ space units from the left-hand domain edge.
In the large core (Fig.~\ref{fig:dnsmultiLC}) a larger $40 \times 40$ square
domain was used, with the step inhomogeneity located $10$ space units from the
left-hand edge, in order to avoid interaction of the spiral wave with the
no-flux domain walls. The grid spacing was $h=0.025$, with corresponding time
step $\triangle t = 9.4 \times 10^{-5}$.
Model parameter values are given later in the text. 


\section{Results}\label{sec:results}

Before presenting our response function calculations, we make a note concerning
incident and reflected angles. As is standard, we define both the angles of
incidence $\thi$ and reflection $\thr$ to be measured from the boundary normal.
In the case of light paths in classical optics, one considers incident angles
only in the range $[0^\circ,90^\circ]$, since, due to symmetry in the
$y$-direction, trajectories at equal angles either side of the normal
correspond to physically identical situations. However, since spirals possess a
chirality, this symmetry is not present and we must consider both incident and
reflected angles in the range $[-90^\circ, 90^\circ]$.

In Sec.~\ref{sec:theory} and the appendix we have implicitly set $\omega > 0$
to correspond to clockwise rotation. We consider spirals of this chirality
only.  Our convention is to define $\thi$ to be positive in the clockwise
direction from the normal and $\thr$ to be positive in the counterclockwise
direction from the normal. That is, incident and reflected angles on
\emph{opposite} sides of the normal have the same sign.

\subsection{Small-core case}\label{sec:scresults}
\begin{figure}[t]
\centering
\includegraphics[width=2.5in]{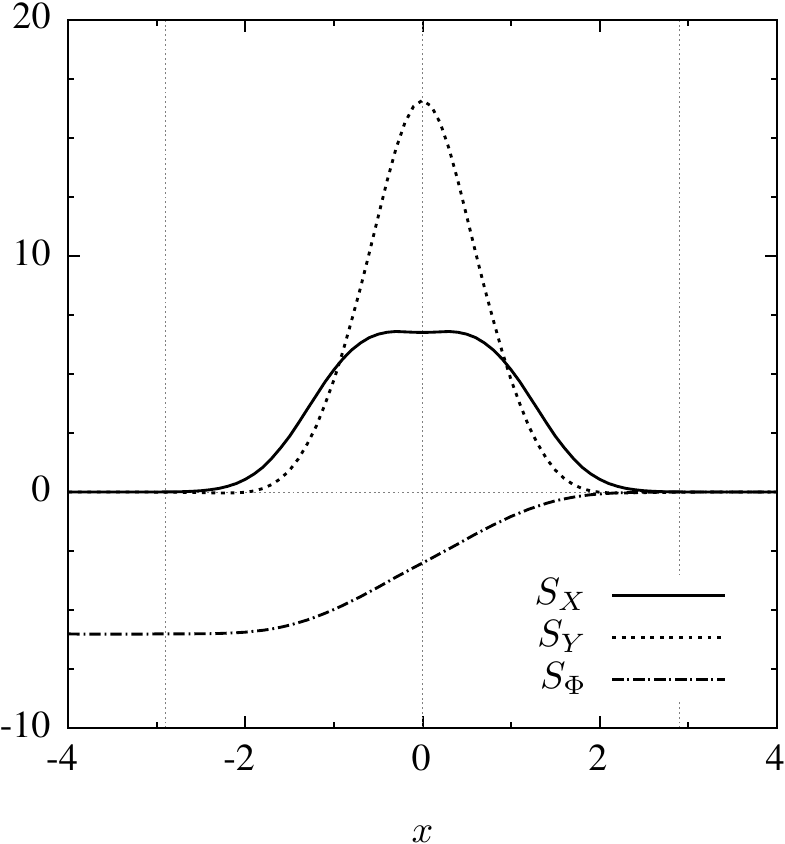}
\caption{$S_X$, $S_Y$, and $S_\Phi$ for a small-core spiral with $a = 0.8$,
$b=0.05$, and $c=0.02$. Also plotted in dotted gray are the vertical lines $x =
\pm 2.9$, which enclose the effective boundary region. [For $|x| > 2.9$,
$S_X(x)$ and $S_Y(x)$ are less than 0.1\% of $S_X(0)$ and $S_Y(0)$
respectively.]}
\label{fig:stepint_sc} 
\end{figure}

Our study begins by considering spiral waves in the small-core region of
parameter space.  We set $a = 0.8$, $b = 0.05$, and $c=0.02$.
Figure~\ref{fig:stepint_sc} shows the step boundary functions $S_X$, $S_Y$, and
$S_\Phi$ for these parameters. 
These curves represent the intrinsic character of the boundary influence. Let
us first consider the effects of this boundary in the absence of resonant
forcing.  The dynamics of the spiral rotation center in this case are governed
simply by the $S_X$ and $S_Y$ curves, scaled by the size of the step:
\begin{equation}
\dot{R} = \epsilon_s S(X)
\end{equation}
where $S\equiv S_X + iS_Y$. We see, as expected, that $S_X$ and $S_Y$ are zero
outside a relatively small neighborhood of $x=0$ and thus spirals outside this
region are unaffected by the step boundary.  Since $S_X(X)$ is positive inside
the boundary region, spirals to the right of $x=0$ are repelled away from the
step.  
\begin{figure}[h]
\centering
\includegraphics[width=2.75in]{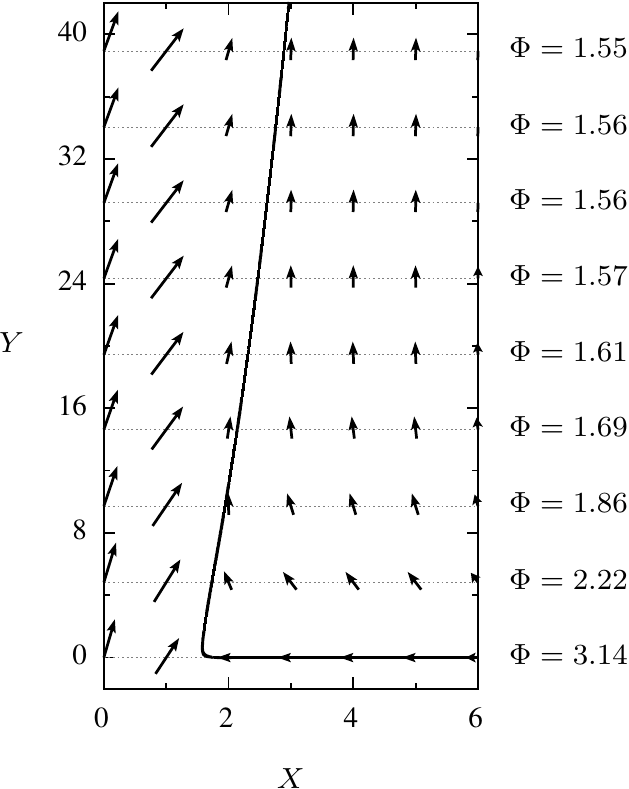}
\caption{Theoretical trajectory of a small-core spiral wave reflection
with $\theta_i=0^\circ$ and $\epsilon_f/\epsilon_s = 1/25$. Initial conditions:
$X_0 = 6$, $Y_0 = 0$, $\Phi_0 = \pi$.  Each horizontal row of vectors plots the
velocity field at the instant at which the spiral center attained the given $Y$.
These vectors depend on $X$ and the phase $\Phi$. The value of $\Phi$ at each
horizontal slice is indicated on the right-hand side. Vector magnitudes have
been scaled nonlinearly for visual clarity. The ratio of the $X:Y$ scales is
$1:4$.}
\label{fig:scvectors} 
\end{figure}
Furthermore, since $S_Y(X)$ is also positive in this region, the
boundary acts to intrinsically drive spirals in the positive $y$-direction.
Note also the antisymmetry of $S_\Phi$. Far to the left of the boundary,
$S_\Phi(X)$ tends to a non-zero (negative in this case) constant. This is
because the spiral's rotation frequency in the left half-plane, with the
perturbed model parameter $b_0-\epsilon_s$, differs from the `natural'
frequency $\omega$ of the unperturbed spiral in the right half-plane.

Now let us add in the effect of periodic forcing. The rotation center in this
case moves according to
\begin{equation}
\dot{R} = \epsilon_s S(X) + \epsilon_f F(\Phi)
\end{equation}
where $F(\Phi) = Ae^{i\Phi}$, from Eq.~\eqref{eq:resforcing}. Thus, the
velocity at each instant is the superposition of the step component and a
vector of fixed magnitude due to the resonant forcing, whose direction is set
by the spiral's phase $\Phi$. Far from the boundary, the velocity is constant,
since $S(X)=0$ and $S_\Phi(X)=0$ for $X\gg0$. Close to the boundary, if the
step perturbation is large enough relative to the resonant forcing
perturbation, the boundary effects dominate and spirals in the positive
half-plane are repelled from the step.  Furthermore, since $S_\Phi(X) < 0$ for
$X \lesssim 2.9$, the forcing component rotates clockwise in time while the
spiral is in the boundary region.

\begin{figure}[ht]
\centering
\includegraphics[width=2.7in]{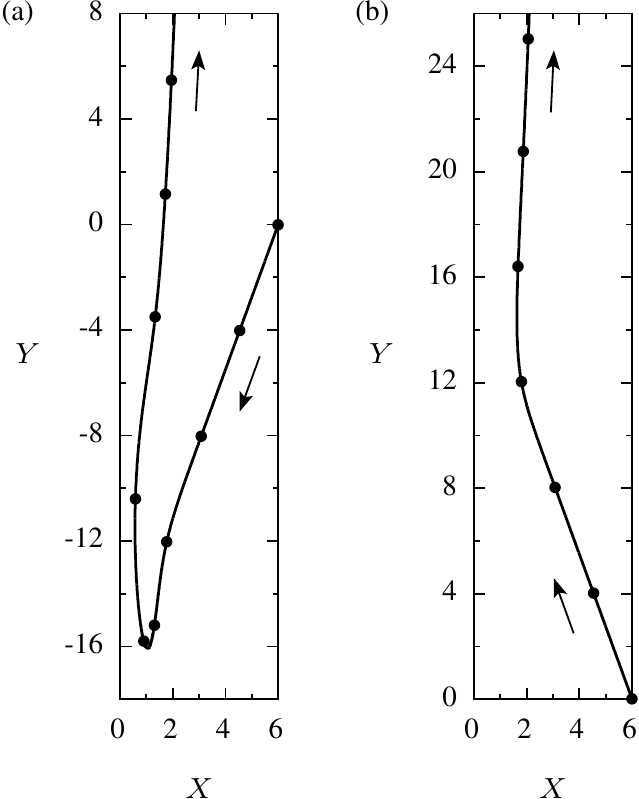}
\caption{Two theoretical trajectories in the small-core regime, initiated at
$X_0=6$, $Y_0=0$.  $\epsilon_f/\epsilon_s = 1/25$. The filled points
plotted along the trajectories are equally spaced in time to indicate drift
speed. Incident angles are (a) $\thi \approx -70^\circ$ and (b) $\thi \approx
+70^\circ$. Both spirals reflect with angle $\thr\approx+88^\circ$. The ratio of
the $X:Y$ scales is $1:1$.}
\label{fig:sc_varioustraj}
\end{figure}
This suggests a mechanism for reflection. Consider a resonantly forced spiral
wave traveling towards the step from the right half-plane. Far from the
boundary, the spiral drifts with constant velocity at some incident angle
$\thi$ (set by initial conditions).  On entering the boundary region, the
spiral is repelled by the inhomogeneity, causing it to slow and preventing it
from passing through $x=0$.  This effect itself does not cause the subsequent
reflection from the boundary.  The motion away from the boundary is rather due
to the $\Phi$ dynamics.  As the spiral approaches the boundary, $\Phi$
decreases bringing about a rotation in the resonant forcing component
$F(\Phi)$.  After a time, this component inevitably rotates around to the
positive $x$-direction and this drives the spiral away from the step.
Consequently, the spiral leaves the boundary at some reflection angle $\thr$,
dictated by the phase on exiting the boundary region.

We see this mechanism at work in Fig.~\ref{fig:scvectors}, which displays a
typical theoretical reflection trajectory in the small-core regime.
(One should note that the lengths of vectors in Fig.~\ref{fig:scvectors} have
been scaled nonlinearly so their directions far from the step are
discernable---the magnitude of the forcing component is comparatively much
weaker than depicted.)
\begin{figure*}[ht]
\centering
\includegraphics[width=6.9in]{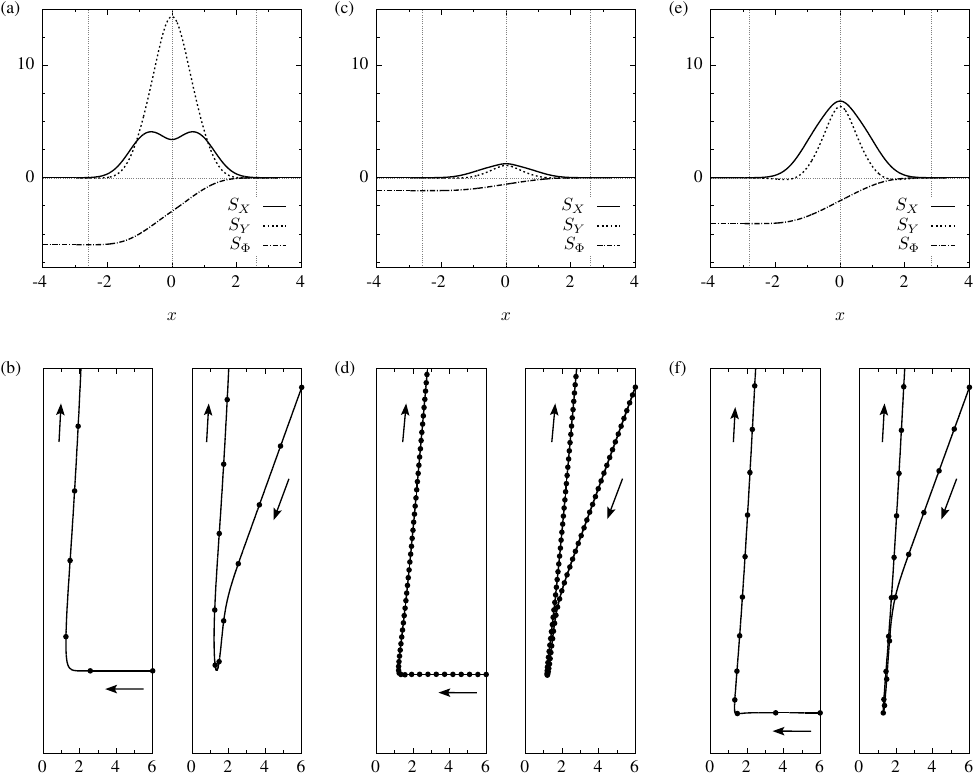}
\caption{$S_X$, $S_Y$, and $S_\Phi$ curves, together with representative
theoretical reflection trajectories for three different small-core
spiral waves.  $\epsilon_f/\epsilon_s = 1/50$. Each pair of reflection
trajectories is plotted below the corresponding boundary curves. The left- and
right-hand trajectories are $\thi \approx 0^\circ$ and $\thi\approx
-70^\circ$, respectively, and include filled points, matched to
the timestep of the corresponding points in Fig.~\ref{fig:sc_varioustraj},
indicating drift speed.
Model parameters: in (a) and (b) $a=0.7$, $b=0.01$; in (c) and (d) $a=0.95$, $b=0.01$;
in (e) and (f) $a=0.95$, $b=0.08$.  In all cases $c=0.02$.  These span a
substantial extent of the small-core regime.}
\label{fig:varstepint_sc}
\end{figure*}
After entering the boundary region, the spiral undergoes a rapid change in
direction and phase and its speed in the $x$-direction slows considerably.  As
the resonant forcing component $F(\Phi)$ (depicted in the rightmost vectors of
Fig.~\ref{fig:scvectors}) rotates with the decreasing phase, its $x$-component
diminishes and consequently the boundary effects push the spiral center further
away from the step. This process is slow and the spiral travels far in the
$y$-direction in this time. Eventually, the evolving phase turns the resonant
drift direction towards the positive half-plane, i.e., $F_X(\Phi)$ changes sign
and becomes positive.  As a result, the spiral center leaves the boundary. The
reflected angle is close to $+90^\circ$, since $S_\Phi(X)$ is very near zero
when this sign change occurs and therefore phase changes only by a small amount
after this. 

We observe that the situation is similar across the full range of incident
angles $\thi \in [-90^\circ,90^\circ]$. Figure~\ref{fig:sc_varioustraj} displays
two theoretical reflection trajectories which approach the boundary at
different angles, either side of the normal, reflecting in the same direction.
Regardless of incident angle, the spiral center may only leave the boundary once
$F(\Phi)$ points away from the step. Each spiral wave reaches this sign change
of $F_X(\Phi)$ in essentially the same state: with $\Phi=\pi/2$ and $X$ close to
the edge of the boundary region.  This is because the $\Phi$ dynamics are
sufficiently slow that the spiral center is pushed almost completely out of the
boundary region by the time that $\Phi=\pi/2$. Therefore each spiral wave changes
direction by only a small amount after this point and reflects with $\thr$ close
to $+90^\circ$. 

It is worth noting that in addition to the invariance of reflection angle,
these theoretical trajectories exhibit qualitative features observed in
numerical simulations. In particular, the nontrivial shape of
Fig.~\ref{fig:sc_varioustraj}(a), the sharp change of direction at the boundary
in Fig.~\ref{fig:scvectors} and the decrease in the closest distance to the
boundary reached by the spiral center as $\thi$ increases. For comparison see
Figs.~3(b), 4(g) and 4(h) in Ref.~\cite{lang13}.

Across the small-core parameter regime, we see that the curves $S_X$, $S_Y$, and
$S_\Phi$ vary in magnitude and shape. However, the qualitative differences in
the theoretical reflection trajectories are only subtle and the
reflection mechanism in each case is the same. Representative curves and
trajectories are plotted in Fig.~\ref{fig:varstepint_sc}.

\begin{figure}[ht]
\centering
\includegraphics[width=2.5in]{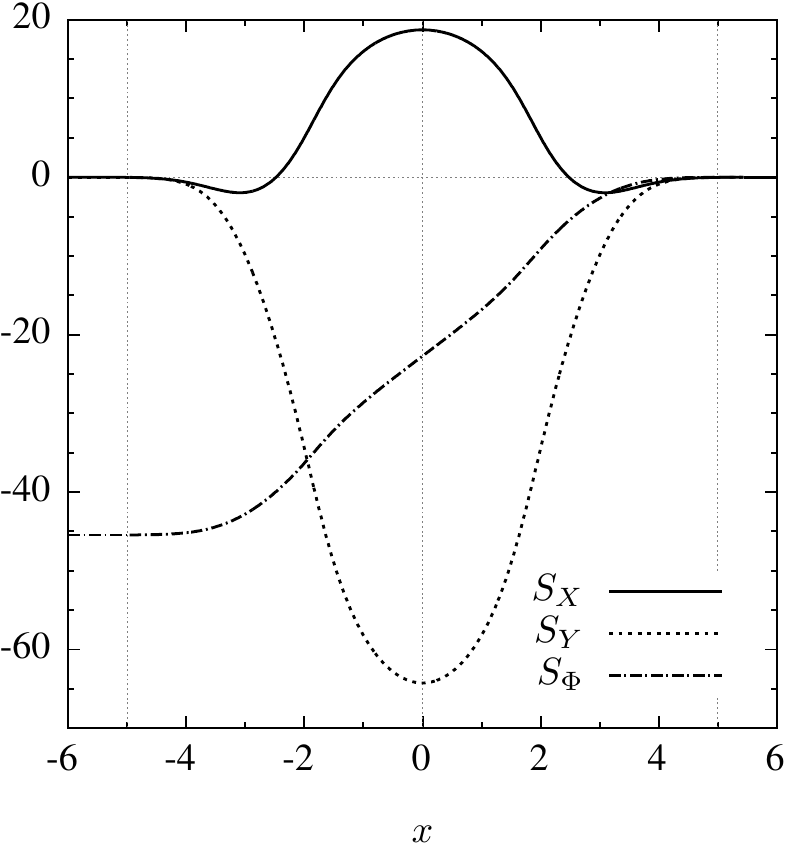}
\caption{$S_X$, $S_Y$ and $S_\Phi$ for a large-core spiral with $a = 0.6$,
$b=0.07$, $c=0.02$. Also plotted in dotted gray are the vertical lines $x = \pm
5.0$, which enclose the effective boundary region. [For $|x| > 5.0$, $|S_X(x)|$
and $|S_Y(x)|$ are less than 0.1\% of $S_X(0)$ and $S_Y(0)$ respectively.]}
\label{fig:stepint_lc} 
\end{figure}
\subsection{Large-core case}\label{sec:lcresults}
We now turn to the large-core case, setting $a=0.6$, $b=0.07$, and $c=0.02$.
As before, we begin by plotting the $x$-dependence of the key functions $S_X$,
$S_Y$, and $S_\Phi$, shown in Fig.~\ref{fig:stepint_lc}.
At first glance these do not appear differ too much from the corresponding
curves in the small core (see Figs.~\ref{fig:stepint_sc}
and~\ref{fig:varstepint_sc}). Nevertheless, there are differences, some of
which are quite important.  The region of boundary influence is wider than in
the small-core, extending to roughly a distance of five space units from the
step inhomogeneity.  This is expected: spiral waves propagate outwards from
their tips, which rotate around a circle of much larger radius in the
large-core.  
Furthermore, $S_X$ has roots within this boundary region, at
$x\approx\pm2.5$.  The root at positive $x$ is attracting (in the absence of
resonant forcing).  Also, the magnitudes of the curves are (pointwise) greater
than those in the small-core case.  For the set of parameters we consider, this
is particularly true for $S_\Phi$.  Finally, notice that $S_Y$ has changed sign
with respect to the small-core case. 
\begin{figure}[ht]
\centering
\includegraphics[width=2.8in]{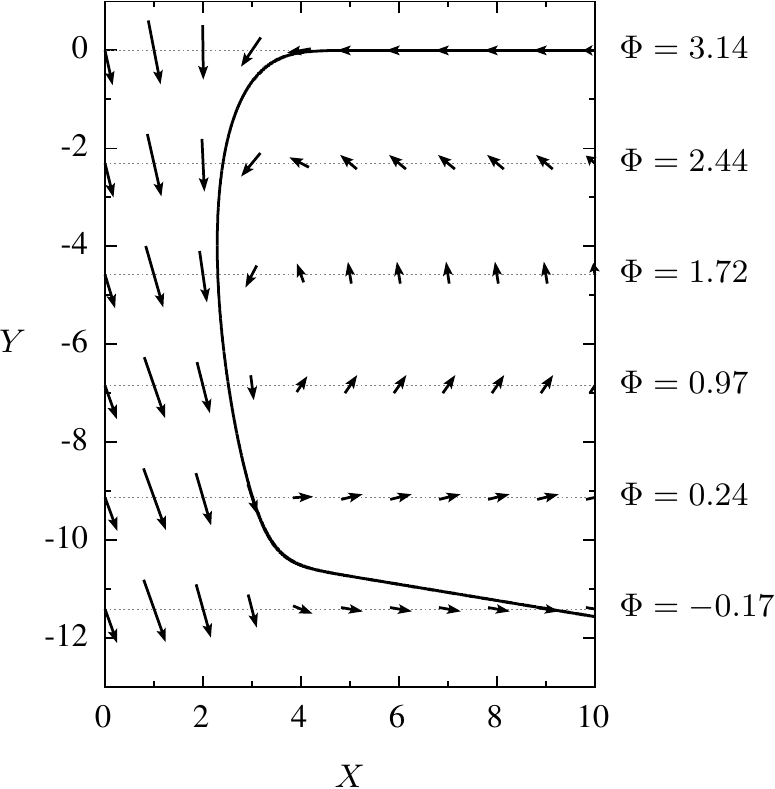}
\caption{Theoretical trajectory of a large-core spiral reflection with
$\theta_i=0^\circ$ and $\epsilon_f/\epsilon_s = 1/87.5$. Initial conditions:
$X_0 = 10$, $Y_0 = 0$, $\Phi_0 = \pi$.  Each horizontal row of vectors plots the
velocity field at the instant at which the spiral center attained the given $Y$.
These vectors depend on $X$ and the phase $\Phi$. The value of $\Phi$ at each
horizontal slice is indicated on the right-hand side. Vector magnitudes have
been scaled nonlinearly for visual clarity. The ratio of the $X:Y$ axes is
$1:1$.}
\label{fig:vectorfield_lc}
\end{figure}

These differences have a significant impact on the character of reflections for
spiral waves in the large-core region. Figure~\ref{fig:vectorfield_lc}
demonstrates a typical theoretical trajectory.
Approaching at $\thi=0^\circ$, the spiral changes direction as it enters the
boundary region as before, but turns to move in the negative rather than the
positive $y$-direction, since $S_Y$ is large and negative inside the boundary
region. While $\pi/2 < \Phi < \pi$, the resonant forcing has negative
$x$-component and the spiral remains near the positive root of $S_X$. Once
$\Phi$ decreases to less than $\pi/2$, the forcing acts to push the spiral away
from the boundary. As it exits, $\Phi$ continues to decrease causing the
resonant forcing direction to turn further clockwise. Finally, the spiral
leaves the boundary at the constant angle dictated by $\Phi=-0.17$ ($\thr
\approx -9.5$ in this case). Qualitatively similar trajectories for low
amplitude resonant forcing in the large core have been observed previously for
Neumann boundary conditions; see Fig.\ 10(c) of Ref.~\cite{lang13}.

The key difference between this large-core case and the small-core
theoretical trajectories in Sec.~\ref{sec:scresults} is the attracting
root of the $S_X$ curve, which importantly occurs within the boundary region.
\begin{figure}[h]
\centering
\includegraphics[width=3.25in]{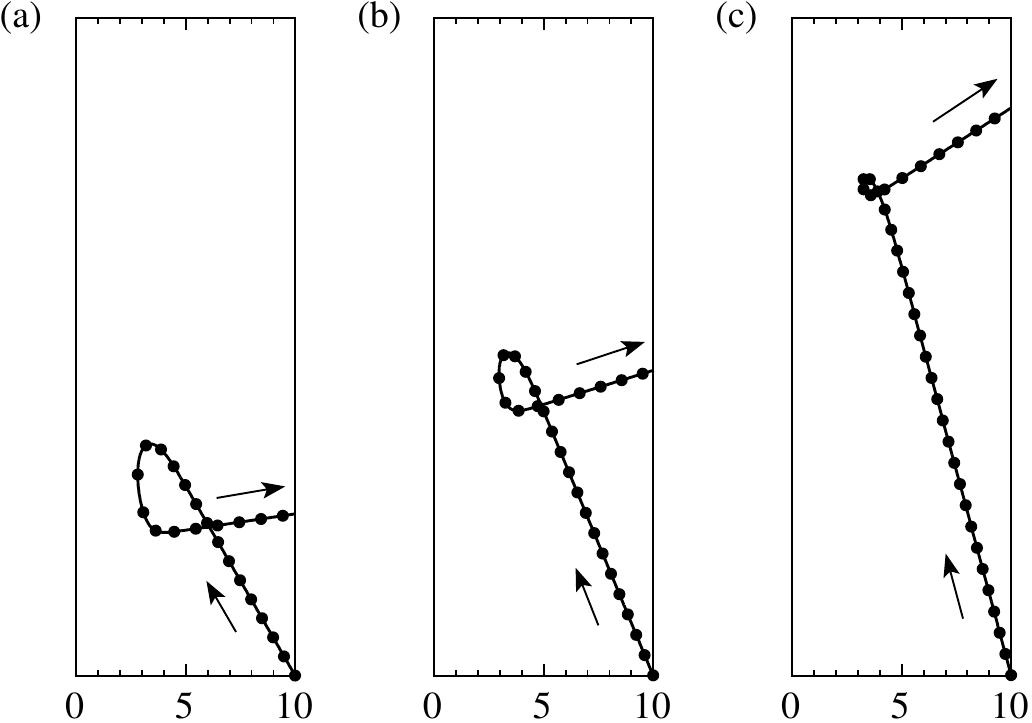}
\caption{Effect of incident angle $\thi$ for a large-core spiral. Various
theoretical trajectories are shown with different initial $\Phi_0$ and
$\epsilon_f/\epsilon_s=1/87.5$. The filled points plotted along the
trajectories are equally spaced in time to indicate drift speed. Incident
angles: (a) $\thi = 60^\circ$, (b)
$\thi = 67.5^\circ$, (c) $\thi = 75^\circ$.}
\label{fig:lc_inceffect}
\end{figure}
While the spiral is in the boundary region, the phase evolves, causing the
resonant forcing component to rotate, just as with small-core spirals. Once
$F_X(\Phi)$ changes sign, the resonant forcing turns to impel the spiral away
from the boundary.
While in the small-core cases this occurs when the spiral center is near to the
end of the boundary region, in the large-core case the spiral remains close to
the attracting root of $S_X$ prior to the sign change.  Since the magnitude of
$S_\Phi$ is non-negligible near the attracting root of $S_X$, $\Phi$ continuous
to evolve, decreasing for some time as the spiral exits the boundary.
Consequently, the final direction of the spiral differs greatly from
$+90^\circ$.

\begin{figure}[ht]
\centering
\includegraphics[width=3.25in]{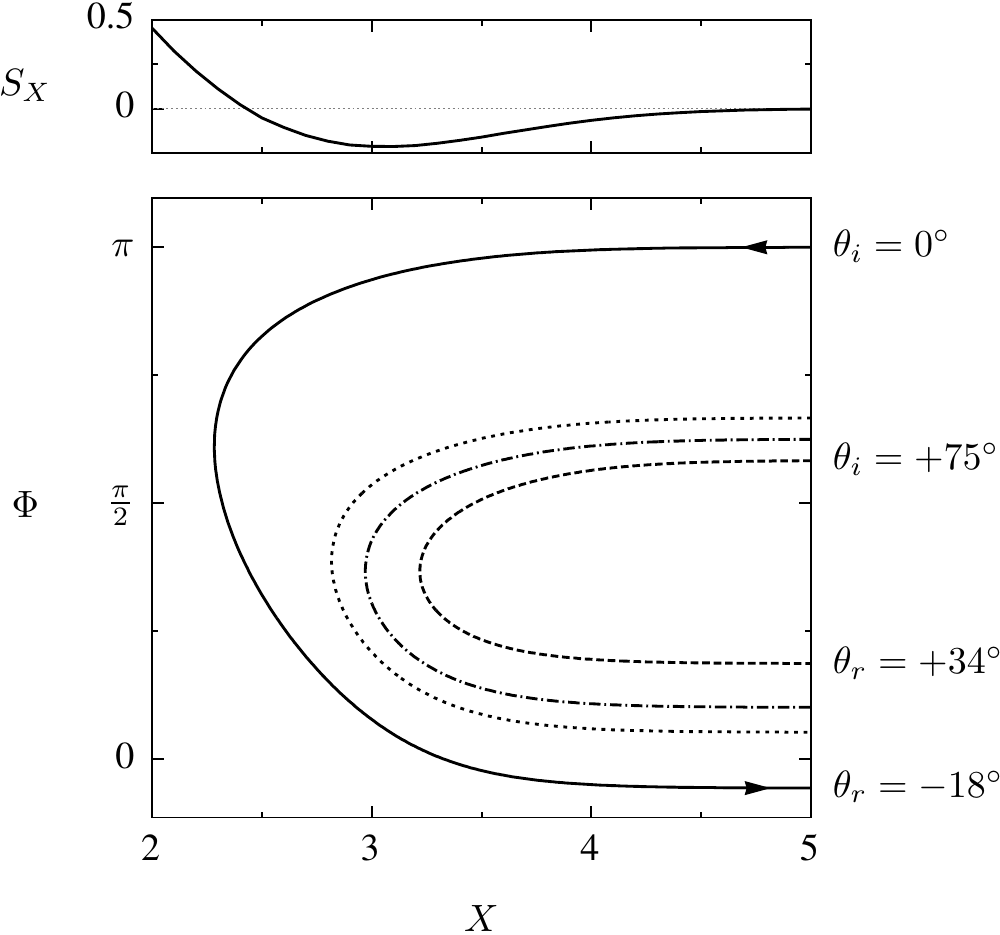}
\caption{Phase dynamics for large-core spirals approaching the boundary with
different incident angles. $\epsilon_f/\epsilon_s=1/87.5$. The top plot shows
the curve $S_X$, for reference.  The bottom plot shows the theoretical
`trajectory' of the spiral phase as the spiral moves in and out of the boundary
region, for various incident angles. Incoming trajectories have $\Phi\in
(\pi/2,3\pi/2)$ and outgoing trajectories have $\Phi\in (-\pi/2,\pi/2)$. The
solid black trajectory corresponds to the reflection in
Fig.~\ref{fig:vectorfield_lc} and the dotted and dashed trajectories correspond
to the reflections in Fig.~\ref{fig:lc_inceffect}.}
\label{fig:XPhitrajs}
\end{figure}

In the large-core regime, we see a notable effect of incident angle on
reflection angle. Using the same parameters, we demonstrate this in
Fig.~\ref{fig:lc_inceffect}.
Spirals approaching the boundary at higher incidence angles have lower initial
phase and consequently reach the sign change of $F_X(\Phi)$ (at $\Phi = \pi/2$)
sooner. Therefore, at high incident angles the sign change occurs much further
from the step than at low incident angles, since $\Phi$ reaches $\pi/2$ before
the spiral center reaches the attracting root of $S_X$. This means these
spirals necessarily leave the boundary region sooner and with a \emph{greater}
$\Phi$, i.e.,\ greater reflected angle.  This can be visualized more clearly
by plotting the trajectory of the phase with respect to the distance from the
boundary, as we have done in Fig.~\ref{fig:XPhitrajs}.

The change in sign of the $S_Y$ curve between the large- and small-core
parameter regimes has no effect on reflection angle, since the dynamics of the
spiral center far from the boundary depends only on $\Phi$ and $X$. 
However, it
is relevant to the overall qualitative shape of trajectories at the boundary.
This difference in sign can be qualitatively explained by referring to
arguments given by Krinsky \emph{et al.}~\cite{Krinsky:1996} for the case of
spiral wave drift in electric fields, which were later studied by Xu \emph{et
al.}~\cite{Xu:2009} for medium inhomogeneities. Drift of the spiral rotation
center may be caused by changes to the radius of the rotation core and also by
changes to the rotation frequency.  In the Barkley model, decreasing the $b$
parameter, as we have done to form the step boundary, \emph{decreases} the core
size and \emph{increases} the rotation frequency. The effect of our step
inhomogeneity on the core radius causes the spiral to drift in the negative
$y$-direction. However, the effect on the rotation frequency causes the spiral
to drift in the positive $y$-direction.  For small-core parameters, the core
radius changes little and the effect of the step boundary on the rotation
frequency dominates. In the large-core parameter region, it is instead the
changes in the core radius which dominate. Therefore the vertical component of
drift due to the boundary changes sign between the two parameter regions.

\begin{figure}[t]
\centering
\includegraphics[width=3.25in]{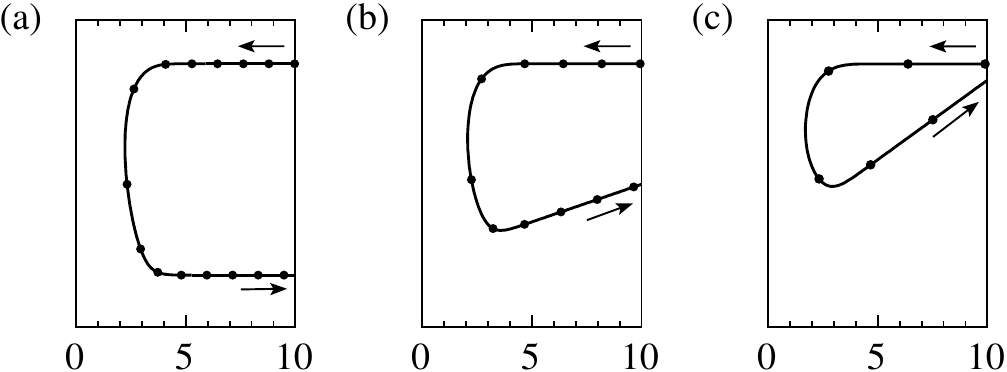}
\caption{Effect of forcing amplitude on large-core spiral waves. Three
theoretical trajectories are shown in order of increasing amplitude and
include filled points, matched to the timestep of the corresponding points in
Fig.~\ref{fig:lc_inceffect}, indicating drift speed.
The perturbation ratio $\epsilon_f/\epsilon_s$ in each case equals (a) $1/75$,
(b) $1/50$, and (c) $1/25$.}
\label{fig:lc_ampeffect}
\end{figure}
We may also consider the effects of altering the ratio $\epsilon_f/\epsilon_s$.
Let us fix $\epsilon_s$ and vary $\epsilon_f$. Higher $\epsilon_f$ corresponds
to higher amplitude resonant forcing, meaning that the drift speed due to
resonant forcing is greater. Figure~\ref{fig:lc_ampeffect} plots some
illustrative theoretical reflection trajectories at different amplitudes.
\begin{figure}[ht]
\centering
\includegraphics[width=3.2in]{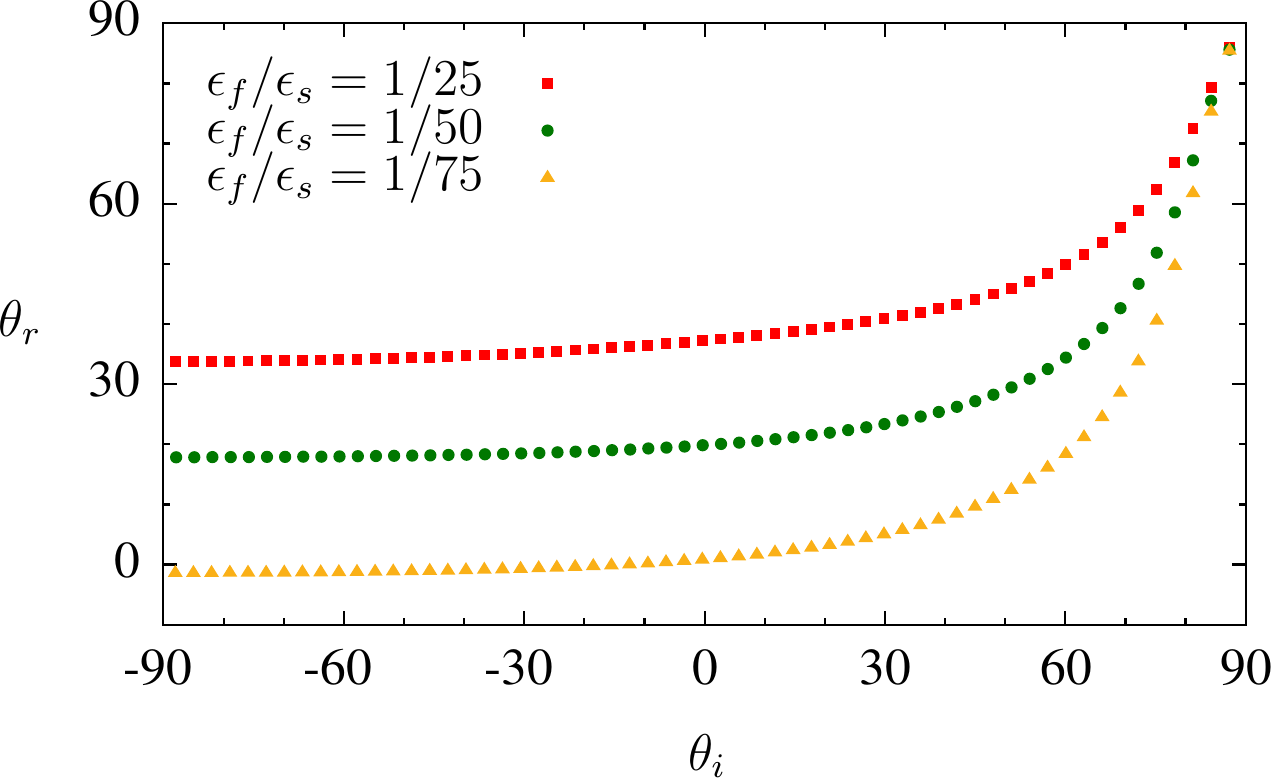}
\caption{(Color online) Reflected angle $\thr$ versus incident angle $\thi$ for large-core
spirals at different forcing amplitudes. The angles were measured from
the theoretical response function trajectories at the given
$\epsilon_f/\epsilon_s$ ratios.}
\label{fig:lc_increfl}
\end{figure}
\begin{figure*}[t]
\centering
\includegraphics[width=6.9in]{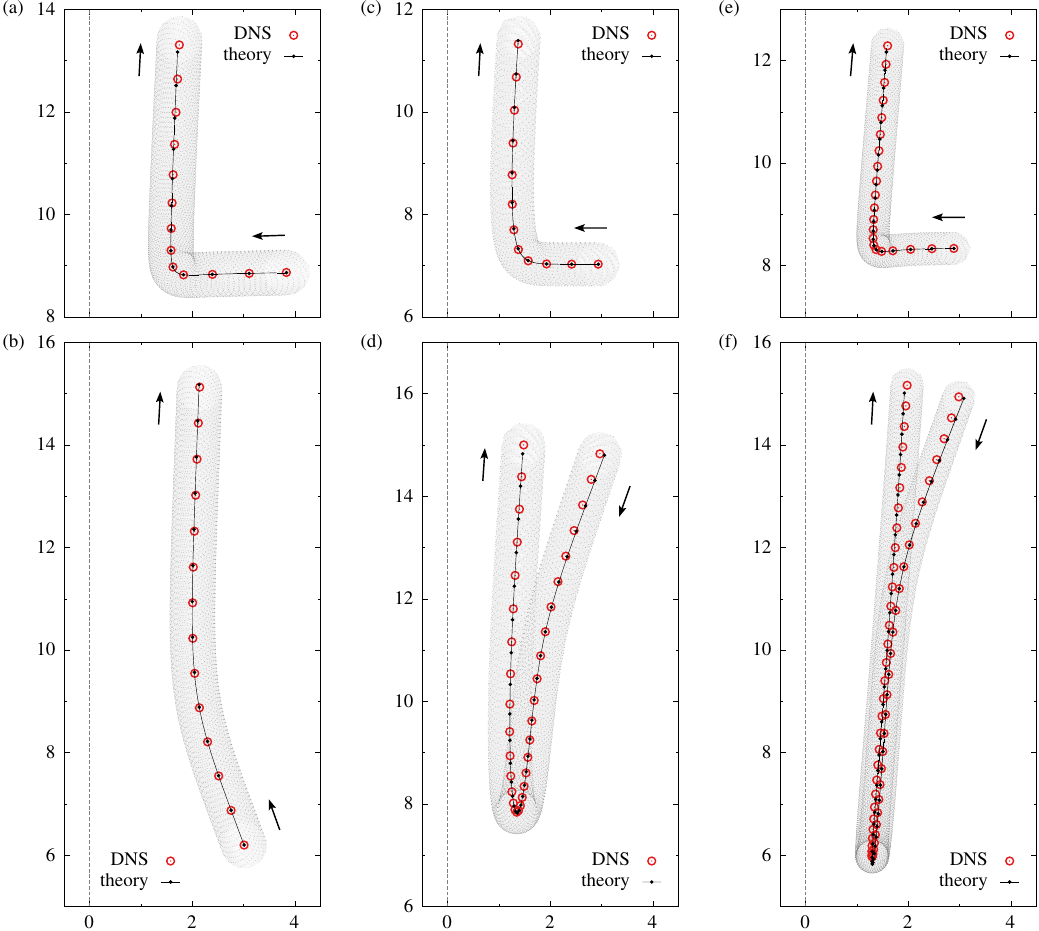}
\caption{(Color online) Comparison between theory and direct numerical simulation
(DNS) of the Barkley model for a variety of parameter values and incident
angles in the small-core regime.
In each case the rotation center of the spiral wave in the DNS is plotted (open
circles) every 30th rotation period. The theoretical trajectories
(curves with solid dots) use an initial condition selected such that they
agree with the DNS trajectory at a point close to the boundary. Solid dots are
separated by a time corresponding to 30 rotation periods. Also shown
are the rotating spiral tip trajectories, dotted in gray, and the step boundary
at $x=0$, dashed in gray.
Each of the three columns corresponds to a different choice of model
parameters broadly spanning the small-core parameter regime. Within each
column two incident angles are shown: one normal and one oblique to the
boundary.
Parameters in (a) and (b): $a = 0.8$, $b=0.05$, $c=0.02$,
$\omega_f=1.850564$, 
$\epsilon_f/\epsilon_s = 4\times 10^{-5}/3.5\times 10^{-3} = 1/87.5$; 
in (c) and (d): $a=0.7$, $b=0.01$, $c=0.02$,
$\omega_f=2.043489$,
$\epsilon_f/\epsilon_s = 4\times 10^{-5}/2\times 10^{-3} = 1/50$;
in (e) and (f): $a = 0.95$, $b=0.08$, $c=0.02$,
$\omega_f=1.768359$,
$\epsilon_f/\epsilon_s = 4\times 10^{-5}/2\times 10^{-3} = 1/50$.}
\label{fig:dnsmultiSC}
\end{figure*}
We see that as resonant forcing amplitude increases, reflected angle increases.
This is because higher amplitude forcing impels spirals with greater drift
speed. Faster spirals leave the boundary more quickly after $F_X(\Phi)$ changes
sign and therefore leave with a greater $\Phi$. (Note that they also approach
closer to the step, which acts to decrease reflection angle, but this effect is
not significant relative to the effect of increased drift speed.)

The combined effects of incidence angle and forcing amplitude are illustrated
in Fig.~\ref{fig:lc_increfl}, where we plot reflected angle $\thr$ versus
incident angle $\thi$ for the three forcing amplitudes used in
Fig.~\ref{fig:lc_ampeffect}.
These theoretical incidence-reflection data are qualitatively close to previously
reported large-core results from direct numerical simulation (albeit with
Neumann boundary conditions): see Fig.\ 9 of Ref.~\cite{lang13}.

\begin{figure*}[ht]
\centering
\includegraphics[width=6.9in]{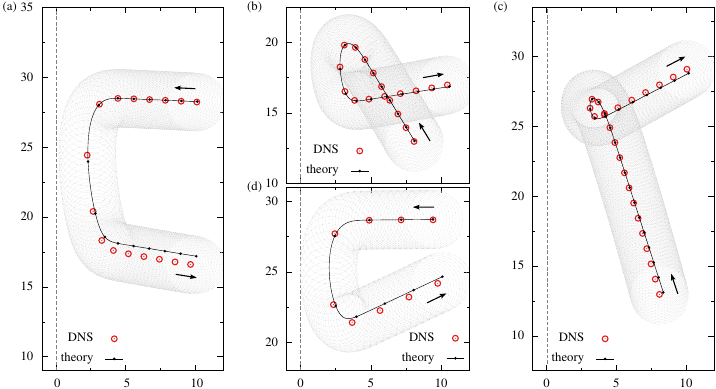}
\caption{(Color online) Comparison between theory and direct numerical simulation
    (DNS) of the Barkley model in the large-core regime verifying the
    theoretical predictions for the role of incident angle and forcing
    amplitude.
Plot (a) shows a simulation of the case explained
theoretically in Fig.~\ref{fig:vectorfield_lc}, in which $\thi\approx 0^\circ$.
Between (a), (b) and (c), the incident angle was varied from $0^\circ$ to
approximately $60^\circ$ and $70^\circ$, respectively, keeping all other
parameters fixed. 
In (d), the resonant forcing perturbation used was twice that of (a),
while the incident angle and all other parameters remained fixed.  
Thus the effects of incident angle and forcing amplitude are seen to agree
with those predicted in Figs.~\ref{fig:lc_inceffect}
and~\ref{fig:lc_ampeffect} and explained theoretically in
Sec.~\ref{sec:lcresults}.
In each case the rotation center of the spiral in the DNS is plotted (open
circles) every 10th rotation period.  The theoretical trajectories
(curves with solid 
dots) use an initial condition selected such that they agree with the DNS
trajectory at a point close to the boundary and are plotted with a time step
(time between successive points) corresponding to 10 rotation periods of
the simulation. Also shown are the rotating spiral tip trajectories, dotted in
gray, and the step boundary at $x=0$, dashed in gray.
Parameters: $a = 0.6$, $b=0.07$, $c=0.02$, and $\omega_f=0.9164372$;
in (a)--(c) $\epsilon_f/\epsilon_s = 4 \times 10^{-5}/3.5 \times 10^{-3} =
1/87.5$; in (d) $\epsilon_f/\epsilon_s = 8 \times 10^{-5}/3.5 \times 10^{-3} =
1/43.75$.
}
\label{fig:dnsmultiLC}
\end{figure*}

\subsection{Comparison with direct numerical simulation}

Figures~\ref{fig:dnsmultiSC} and~\ref{fig:dnsmultiLC} show comparisons between
the reflection trajectory predicted by our response function calculations and
results from direct numerical simulation (DNS) of the full Barkley model PDEs
using the same parameters.
A thorough study of the \emph{numerical convergence} of the asymptotic theory
in the separate cases of resonant parameter forcing and step inhomogeneity has
previously been conducted~\cite{Biktasheva:2010} and consequently we do not 
repeat such a study here.
Instead, the cases presented have been chosen to demonstrate various phenomena
predicted theoretically in the preceding sections. Excellent agreement is seen
between theory and full DNS of spiral waves over a broad range of parameters
and conditions.

In the small-core cases, Fig.~\ref{fig:dnsmultiSC}, the spiral wave drift
direction, drift speed, and point of closest approach to the boundary are in
very close correspondence with theoretical predictions. Note that speed is
gauged from the distance traveled between successive points (open circle for
DNS and filled circles for theory). Most of the (very small) differences between
DNS and theory arise in the vicinity of the boundary where the effects of both
perturbations are felt. Since points are plotted at fixed time intervals over
the full trajectory, small speed differences can nevertheless give rise to an
accumulated shift between points from DNS and theory.  The most striking feature
in the small-core regime is the correct theoretical prediction at large negative
incident angles: Figs.~\ref{fig:dnsmultiSC}(d) and~\ref{fig:dnsmultiSC}(f).
Theory correctly predicts that the spiral center first moves downward near the
boundary for a large number of rotation periods before turning, moving upward,
and slowly leaving the boundary.

In the case of large-core spiral waves, Fig.~\ref{fig:dnsmultiLC}, the
considerable variation in the reflected angle predicted by theory is seen to
hold in the full DNS.  In particular, for fixed parameter values, as the
incident angle is changed from near zero, Fig.~\ref{fig:dnsmultiLC}(a), to large
positive angles, Figs.~\ref{fig:dnsmultiLC}(b) and~\ref{fig:dnsmultiLC}(c), the
drift trajectory spends less time in the vicinity of the boundary and develops a
loop as the reflected angle changes from negative (moving down and to the right
in the figure) to positive (moving up and to the right). (See also for
comparison Fig.~\ref{fig:lc_inceffect}.)  Furthermore, as the forcing amplitude
is increased for otherwise fixed conditions [Fig.~\ref{fig:dnsmultiLC}(a) and
Fig.~\ref{fig:dnsmultiLC}(d)], the time at the boundary decreases and the
reflection angle increases.  (See also for comparison
Fig.~\ref{fig:lc_ampeffect}.)

The agreement between asymptotics and DNS is not quite as good in the
large-core results, Fig.~\ref{fig:dnsmultiLC}, as in the small-core results,
Fig.~\ref{fig:dnsmultiSC}. The main visible difference between theory and DNS
in the large core regime is the point at which the spiral center leaves the boundary. Other
features, such as the reflected angle and the point of closest approach are
predicted well. Discrepancies between theory and DNS 
are due to slight frequency mismatches. Large-core spiral waves are
particularly susceptible to this as their rotation frequencies and tip orbits vary
rapidly with parameters~\cite{Winfree:1991}.
In the DNS there is a shift from the unperturbed rotation frequency $\omega$ (as
calculated to high accuracy by \textsc{DXSpiral}) due to small but finite
spatial discretization errors, as well as weak nonlinear effects at
finite perturbation strength.
As the perturbation magnitudes and the computational grid spacing tend to
zero, the theoretical and DNS trajectories do converge~\cite{Biktasheva:2010}.


\section{Discussion}\label{sec:discussion}

We have applied the theory of response functions to the reflection of spiral
wave trajectories from boundaries. Via numerical computation of response
functions, we have studied reflections in the asymptotic limit of slow drift
and weak boundary effects.  
In this limit the approach is quantitatively accurate, as we have
  demonstrated for a variety of cases by comparing direct simulations of
  spiral waves in a full reaction-diffusion model with the theoretical
  predictions from response functions. 
  However, the main value of the response function
  approach is the {\em qualitative} understanding it brings to how
  interactions with a boundary lead to different types of reflections in
  various situations.
Several of the most
significant features of spiral wave reflections, previously observed in
simulations at higher drift speeds and greater step
inhomogeneities~\cite{lang13}, are 
nevertheless captured qualitatively by the asymptotic analysis.
Consequently, we have been able to understand the essential causes of many
interesting aspects of spiral wave reflections.

As stated in the Introduction, the primary characteristic of spiral wave
reflections is that across a wide range of model parameters, the reflected angle
is approximately constant for large ranges of incident angle.  This reflection
angle `plateau' is present in the response function results in both small- and
large-core cases.  In the small-core case, it was previously demonstrated
numerically that the value of this constant angle increases toward $\thr =
+90^\circ$ as the resonant drift velocity decreases~\cite{lang13}. 
Our asymptotic results reveal the limiting case of this trend,
yielding only reflected angles very close to $\thr = +90^\circ$ and we
have shown exactly why the reflected angle is essentially constant across a wide
range of parameter space. 

Another significant feature observed in prior numerical simulations of
reflections is that, unlike the small-core case, for large-core spirals the
reflected angle \emph{increases} with increasing drift
velocity~\cite{lang13}.  This effect is clearly present in the asymptotics
(Figs.~\ref{fig:lc_ampeffect} and~\ref{fig:lc_increfl}) and in the comparison with DNS
[Figs.~\ref{fig:dnsmultiLC}(a) and~\ref{fig:dnsmultiLC}(d)].
The qualitative form of the reflection angle data in Fig.~\ref{fig:lc_increfl}---a plateau for
negative $\thi$, then monotonically increasing at high $\thi$---is familiar to
all previous numerical results and emerges naturally from the response function model
by considering Fig.~\ref{fig:XPhitrajs}. Furthermore, general
consideration of the differences between small- and large-core spiral waves at
the asymptotic level has led to explanations of the diversity of behaviors
between the two cases. Finally, we note that the non-trivial shape, closest
boundary approach distance, and relative drift speeds that are obtained
and explained via the response function analysis are all observed
qualitatively beyond the asymptotic limit, in both the small- and large-core
cases~\cite{lang13}.

The work presented in this paper fits comfortably with that which is already
known about spiral wave reflections. Biktashev and Holden~\cite{bikt93}
recognized many years ago that reflections are caused by small deviations from
the natural rotation frequency on close approach to the boundary, which alter
the direction of drift. They proposed asymptotic equations of motion for the
rotation center and phase, positing that the boundary effects (corresponding to
$S_X$, $S_Y$, and $S_\Phi$ in our notation) decay exponentially with distance
from the boundary.  These simple assumptions ably capture the overriding
feature of spiral wave reflections---large ranges of approximately constant
reflected angle---but beyond that the predictive qualities of the model are
limited.  Our application of response functions to the reflection problem can
be viewed as an extension of their efforts, removing the phenomenology for the
case of a step boundary and allowing the boundary effects to be calculated
accurately for any spiral wave.  This extra information yields a much more
detailed picture of the reflection dynamics, capturing the behavior near to
the boundary as well as far from it and producing qualitatively meaningful
reflection trajectories across a wide range of parameters. Furthermore, we have
calculated response functions in the large-core regime, which was not
considered by Biktashev and Holden. Here, we observe that the repulsive effect
on spirals' velocity normal to boundary ($S_X$) decays more rapidly than the
effect on the phase ($S_\Phi$)---a finding which accounts for the differences
between small- and large-core reflection angle results. This could not have been
captured by the original Biktashev-Holden theory which for simplicity assumed
that all boundary effects decay with respect to the same length scale.

Beyond the features of spiral wave reflections considered here, there are
phenomena outside the asymptotic limit of small perturbations that are not
predicted by the linear order response function approach. In the small-core
regime, a wider range of reflection angles are observed at higher forcing
amplitudes than is captured by the asymptotic analysis.  In the large-core
regime, there exist so-called `glancing' and `binding' trajectories in which
spiral waves respectively become temporarily and permanently attached to the
boundary~\cite{lang13}. It would be desirable to address these phenomena
theoretically---particularly the attachment behaviors which are especially at
odds with what we have seen in the asymptotics. 

One potential approach could be to use a kinematic model, similar to the one
introduced by Di \emph{et al.}\ in Ref.~\cite{Di:2003}.  The principle idea is
to split the motion of the spiral tip into angular and radial components, which
depend on the tip rotation radius $R_c$ and rotation period $T$. The dependence
of $R_c$ and $T$ on the medium parameters (or on some external perturbation) may
be determined empirically by direct simulation and thus used to model drift in a
given scenario.  Recent papers have employed this method to reproduce the tip
dynamics of small- and large-core spirals in the presence of a step
inhomogeneity~\cite{Xu:2009} and under periodic forcing of
excitability~\cite{Xu:2012}. This suggests that a similar approach could be used
to model spiral wave reflections. It remains to be seen whether, given suitable
modeling assumptions, predictive power outside the limit of small perturbations
could be obtained.


\begin{acknowledgments}
We thank V.~N.~Biktashev for the useful discussion. Development
of the DXSpiral software was supported by EPSRC Grants No.\ EP/D074746/1 and
No.\ EP/D074789/1.  Computing facilities were provided by the Centre for Scientific
Computing of the University of Warwick with support from the Science Research
Investment Fund.
\end{acknowledgments}

\appendix*
\section{Response function theory derivations}
In this appendix we present the derivation of the response function inner
products that make up the differential equations in
Eqs.~\eqref{eq:Xdot},~\eqref{eq:Ydot} and~\eqref{eq:Phidot}. 

The perturbations we have considered above are small temporal and spatial
variations in the medium parameters. Denoting the parameter as $p$, we take its
dependence on $(\vect{x},t)$ to be of the form  $p(\vect{x},t) = p_0 + \epsilon
p_1(\vect{x},t)$ for some constants $p_0$ and $0 < \epsilon \ll 1$.  Taylor
expansion of Eq.~\eqref{eq:reactdiff} to first order in $\epsilon$ establishes
that parameter variations of this form may be considered as additive
perturbations to the reaction diffusion system,
\begin{equation}
\partial_t \vect{u} = \vect{D} \nabla^2 \vect{u} + \vect{f}(\vect{u},p_0) + 
\epsilon \vect{h}(\vect{u},\vect{x},t)
\label{eq:reactdiff_perturb}
\end{equation}
where $\vect{h}(\vect{u},\vect{x},t) =
\partial_p\vect{f}(\vect{u},p_0)p_1(\vect{x},t)$. While we could instead
perturb the PDE fields directly, parameter variation is preferred since it is
directly analogous to the way in which experiments on excitable media are often
conducted~\cite{Agladze:1987,Steinbock:1993,Zykov:1994,Kantrasiri:2005}.

\subsection{Resonant forcing} Sinusoidal variation of a parameter at the
natural frequency $\omega$ induces resonant drift. Consider $p$ varying as
$p(t) = p_0 + \epsilon_f\cos(\omega (t-t_0))$, where $t_0$ is some initial time
whose role will become apparent below.  Then the perturbation $\vect{h}_f$, in
the form depicted in Eq.~\eqref{eq:reactdiff_perturb}, is
$\vect{h}_f(\vect{u},t) = \partial_p\vect{f}(\vect{u},p_0)\cos(\omega
(t-t_0))$. 

To derive the dynamical equations for $\Phi$ and $R$, we must perform the
integrations in Eqs.~\eqref{eq:Phidot_general} and~\eqref{eq:Rdot_general}.
Note that since the sinusoidal term does not depend on space:
\begin{equation}
\langle \vect{W}_n, \vect{h}_f \rangle = \cos(\omega (t-t_0))\langle
\vect{W}_n, \partial_p \vect{f}(\vect{u},p_0)\rangle
\end{equation}
for $n=0,1$. Furthermore, both $\vect{W}_n$ and $\partial_p\vect{f}$
depend on time only via their dependence on the wave field $\vect{u}$. Since
$\vect{u}$ is stationary in a reference frame centered at $R$ and rotating with
frequency $\omega$, the inner products $\langle\vect{W}_n,
\partial_p\vect{f}\rangle$ are time independent. Therefore, we have
\begin{equation}
\int_{t-T/2}^{t+T/2} \langle \vect{W}_0,
\vect{h}_f \rangle d\tau = 0
\label{eq:W_0doth_f}
\end{equation}
and
\begin{equation}
\int_{t-T/2}^{t+T/2} e^{i(\Phi-\omega\tau)} \langle \vect{W}_1, \vect{h}_f
\rangle d\tau = \frac{1}{2}T e^{i(\Phi-\omega t_0)} \langle \vect{W}_1, \partial_p
\vect{f}\rangle.
\label{eq:W_1doth_f}
\end{equation}
We set the initial forcing time $t_0$ such that $-\omega t_0 + \arg\langle
\vect{W}_1, \partial_p \vect{f} \rangle = 0$. Therefore the equations of motion
for a sinusoidally forced spiral are, due to
Eqs.~\eqref{eq:Phidot_general},~\eqref{eq:Rdot_general},~\eqref{eq:W_0doth_f},
and~\eqref{eq:W_1doth_f}:
\begin{equation}
\dot{\Phi} = 0, \quad \dot{R} = \epsilon_f Ae^{i\Phi} =
\epsilon_f F(\Phi),
\end{equation}
where $A(\vect{u},p_0) := \frac{1}{2}|\langle \vect{W}_1, \partial_p
\vect{f}(\vect{u}, p_0) \rangle|$ is a real constant with respect to space and
time for a given model and set of parameters. We can thus unambiguously
identify the phase variable $\Phi$ with the direction of drift due to resonant
forcing and it is for this reason that $t_0$ was introduced.

\subsection{Step boundary} 
The step boundary is a step inhomogeneity in a medium parameter that for
convenience we locate at $x=0$. Therefore, the parameter $p$ varies in space as
$p(x) = p_0 + \epsilon_s (H(x) - 1)$, where $H$ is the Heaviside step function.
The perturbation $\vect{h}_s$ is thus $\vect{h}_s(\vect{u},\vect{x}) =
\partial_p\vect{f}(\vect{u},p_0) (H(x)-1)$.

The integrals in Eqs.~\eqref{eq:Phidot_general} and~\eqref{eq:Rdot_general} are
considered here in a co-ordinate system that rotates with the spiral wave at
its natural frequency and is centered at
$R$~\cite{bikt95,bikta03,Biktasheva:2010}.  Let $(\rho,\vartheta)$ be polar
co-ordinates centered at $R$. Then define the rotating angular co-ordinate
$\theta = \vartheta + \phi(t)$, where $\phi(t) := \omega t - \Phi(t)$ is the
angle that the spiral turns through in time $t$. The co-ordinates
$(\rho,\theta,\phi)$ define a frame in which the spiral wave $\vect{U}$ [see
Eq.~\eqref{eq:rigid}] and its response functions $\vect{W}_{0}$ and
$\vect{W}_{1}$ are constant.

In this frame the time-averaging integration in Eqs.~\eqref{eq:Phidot_general}
and~\eqref{eq:Rdot_general} becomes averaging over $\phi$.  [Note that since
the perturbation $\vect{h}_s$ does not depend on time this averaging need not
be centered about $\phi(t)$ and hence we take the range of integration to be
simply $[0,2\pi]$.]
We obtain
\begin{align}
\frac{1}{T}\int_{t-T/2}^{t+T/2} e^{in(\Phi-\omega t)} &\langle \vect{W}_n,
\vect{h}_s \rangle d\tau = \nonumber\\ 
\frac{1}{2\pi}\int_0^{2\pi}e^{-in\phi} &\int_0^{2\pi}\int_0^\infty
w_n(\rho,\theta) \tilde{p}_1(\rho,\theta,\phi) \rho d\rho d\theta d\phi
\label{eq:stepcompute_general}
\end{align}
for $n=0,1$, where $\tilde{p}_1$ represents the spatial variation of $p$
written in the co-rotating frame, which is
\begin{equation}
\tilde{p}_1(\rho,\theta,\phi) = H(X + \rho \cos(\theta - \phi)) - 1,
\end{equation}
and we have made use of the shorthand $w_n := [ \vect{W}_n(\rho,\theta) ]^*
\cdot \partial_p\vect{f}(\vect{U},p_0)$. 

We can compute the integral over $\phi$ explicitly. Changing the co-ordinate to
$\vartheta$ and rescaling the step function, we have
\begin{align}
\frac{1}{2\pi}\int_0^{2\pi} &e^{-in\phi} \tilde{p}_1(\rho,\theta,\phi) d\phi
= \nonumber\\
&\frac{1}{2\pi} e^{-in\theta} \int_0^{2\pi} e^{in\vartheta}
(H(X/\rho+\cos(\vartheta))-1) d\vartheta.
\end{align}
As discussed in the main text, we see that the integral depends on the distance
of the spiral center to the step inhomogeneity.  There are three cases to
consider:
\begin{enumerate}
\item $|X| > \rho$ and $X > 0 \implies H(X/\rho + \cos(\vartheta))-1 = 0$
\item $|X| > \rho$ and $X < 0 \implies H(X/\rho + \cos(\vartheta))-1 = -1$
\item $|X| < \rho$, in which case $H(X/\rho + \cos(\vartheta))-1 = -1$ if
$\vartheta \in [-\pi, -\arccos(-X/\rho)] \cup [\arccos(-X/\rho),\pi]$  and
is zero otherwise.
\end{enumerate}
For the case $n=0$, i.e.,\ the $\Phi$ dynamics, we therefore have
\begin{equation}
\frac{1}{2\pi}\int_0^{2\pi} \tilde{p}_1 d\phi = 
	\begin{cases}
	H(X) - 1 & \text{if } \rho < |X| \\
	\frac{1}{\pi}\arccos(-X/\rho) - 1 & \text{if } \rho > |X|
	\end{cases}
\label{eq:phiint_n=0}
\end{equation}
and for the case $n=1$, i.e.,\ the $R$ dynamics, after some work one obtains
\begin{equation}
\frac{1}{2\pi}\int_0^{2\pi}e^{-i\phi} \tilde{p}_1 d\phi =
	\begin{cases}
	0 & \text{if } \rho < |X| \\
	\frac{1}{\pi\rho}e^{-i\theta}\sqrt{\rho^2 - X^2} & \text{if } \rho > |X|
	\end{cases}.
\label{eq:phiint_n=1}
\end{equation}

Combining the results in Eqs.~\eqref{eq:phiint_n=0} and~\eqref{eq:phiint_n=1}
with Eqs.~\eqref{eq:stepcompute_general} and~\eqref{eq:Phidot_general} we see
that the dynamics for a spiral wave interacting with a step boundary are of the
form 
\begin{equation}
\dot{\Phi} = \epsilon_s S_\Phi(X), \quad 
\dot{R} = \epsilon_s S(X),
\end{equation}
where 
\begin{align}
S_\Phi(&X) = \int_0^{2\pi}\int_0^{|X|}w_0(\rho,\theta)(H(X) - 1)\rho~d\rho
d\theta \nonumber\\
&+ \int_0^{2\pi}\int_{|X|}^{\infty}
w_0(\rho,\theta)\left[\frac{1}{\pi}\arccos(-X/\rho) - 1\right]\rho d\rho
d\theta 
\label{eq:phidot_step}
\end{align}
and 
\begin{equation}
S(X) = \frac{1}{\pi}\int_0^{2\pi}\int_{|X|}^\infty
w_1(\rho,\theta)e^{-i\theta}\sqrt{\rho^2-X^2}~d\rho d\theta .
\label{eq:rdot_step}
\end{equation}
As argued in Sec.~\ref{sec:theory}, the asymptotics for the forcing and step
perturbations linearly superpose, providing the full picture of the dynamics of
a resonantly forced spiral waves interacting with a step boundary. This is
displayed in Eqs.~\eqref{eq:Xdot},~\eqref{eq:Ydot}, and~\eqref{eq:Phidot} with
the $R$ dynamics separated into $X$ and $Y$ components: $S_X := \re(S)$, $S_Y
:= \im(S)$, $F_X := \re(F)$, and $F_Y := \im(F)$.


%


\end{document}